\documentclass[onecolumn,showpacs,preprintnumbers,bibnotes,nofootinbib]{revtex4}
\usepackage{amsmath}
\usepackage{graphicx}
\usepackage{dcolumn}
\usepackage{bm}

\input{tcilatex}

\begin{document}

\title{Entanglement and interference between different degrees of freedom of
photons states}
\author{F. W. Sun,$^{1}\footnote{
Present address: Optical Nanostructures Laboratory, Columbia University, New
York, NY 10027; fwsun@mail.ustc.edu.cn;}$ B. H. Liu,$^{1}$ Y. F. Huang,$^{1}$
Y. S. Zhang,$^{1}$ Z. Y. Ou,$^{1,2}$ and G. C. Guo$^{1}$}
\affiliation{$^1$Key Laboratory of Quantum Information, University of Science and
Technology of China, \\
CAS, Hefei, 230026, the People's Republic of China \\
$^2$Department of Physics, Indiana University-Purdue University
Indianapolis, 402 N. Blackford Street, Indianapolis, IN 46202}
\date{\today }

\begin{abstract}
In this paper, photonic entanglement and interference are described and
analyzed with the language of quantum information process. Correspondingly,
a photon state involving several degrees of freedom is represented in a new
expression based on the permutation symmetry of bosons. In this expression,
each degree of freedom of a single photon is regarded as a qubit and
operations on photons as qubit gates. The two-photon Hong-Ou-Mandel
interference is well interpreted with it. Moreover, the analysis reveals the
entanglement between different degrees of freedom in a four-photon state
from parametric down conversion, even if there is no entanglement between
them in the two-photon state. The entanglement will decrease the state
purity and photon interference visibility in the experiments on a
four-photon polarization state.
\end{abstract}

\maketitle

\section{Introduction}

Photons have been widely applied to quantum information process (QIP), and
photon interference is the kernel of these quantum operations. The optical
part of the experimental quantum key distribution with phase modulation is
the single photon interference. The two-photon Hong-Ou-Mandel interference
(HOMI) \cite{Hong} has been the heart of linear optical quantum computation
\cite{Kok}. Quantum state preparation \cite{Ou88}, manipulation \cite%
{Kiesel,Brien} and measurement \cite{Braunstein,Mattle} can be well realized
with HOMI. Based on these, quantum teleportation was successfully
demonstrated in experiment \cite{Bouwmeester}. Besides, HOMI is well applied
to the purification of quantum state \cite{Ricci0} and the analysis of
photon resource properties \cite{Santori}. It can be used in demonstrations
of quantum cloning and NOT gate \cite{Ricci,Irvine}. Moreover, the
multi-photon interference has been used in the demonstration of the
multi-photon de Broglie wavelength and high resolution quantum measurement
\cite{Fonseca,D'Angelo,Walther,mitchell,sun1}.

In the above QIP protocols, the degree of freedom (DOF) of polarization is
the main role. However, photons involving other DOF are also investigated.
For example, spatial transverse DOF and time-energy DOF can form a
high-dimensional system and also have wide applications in QIP \cite%
{Vaziri,Riedmatten}. High-dimensional entanglement with spatial transverse
DOF has been observed \cite{Mair} experimentally. The DOF of time energy has
been well applied to the long distance quantum communication \cite{Marcikic}%
. Recently, there appears a hyper-entangled two-photon state which is
entangled in several DOF. States entangled in polarization, spatial mode and
time energy \cite{Barreiro}, and in polarization and momentum (path) \cite%
{Cinelli,Barbieri} have been successfully generated and well discussed in
experiment. However, they do not involve the entanglement between different
DOF.

Usually, photon pairs are produced in the nonlinear optical processes of
parametric down conversion (PDC). With different types of phase-matching
conditions, experimenters can get photon states entangled in different DOF.
However, the entanglement between different DOF is ignored when one of the
DOF is investigated emphatically. There could be two reasons for legally
ignoring this. One reason is that different types of filters are used in
experiments. For the polarization state, narrow band interference filter and
single-mode fiber act as the filters on frequency and spatial DOF,
respectively. These filters erase the possible correlations that may exist
between polarization DOF and frequency or spatial DOF. Thus, the
polarization DOF is purified. The other reason is that there may be no
correlations between these DOF. For the two-photon state from PDC, such as
the hyper-entangled state \cite{Barreiro,Cinelli,Barbieri}, there is no
evidence that the correlations exist between those DOF. However, for the
multi-photon states from PDC, there will be entanglement between different
DOF even when no entanglement is in the two-photon state from the lower
order PDC. The entanglement will decrease the state purity and photon
interference visibility if the correlated modes are neglected illegally. It
is also the reason for decoherence. Several experiments discussed this kind
of decoherence \cite{Tsujino,Eisenberg}. However, the full description of
the entanglement and decoherence is seldom introduced when the photon states
contain several DOF. It is more complicated for the multi-DOF multi-photon
state.

In this paper, we will give the demonstration of photon interference on
different DOF with the language of QIP. A photon state containing several
DOF is rewritten in an expression based on the permutation symmetry of
bosonic particles. In this state description and interference demonstration,
each degree of freedom of a single photon is regarded as a qubit and an
operation on photons is regarded as a single-qubit gate or two-qubit gate.
As an application, two-photon HOMI is well re-interpreted. Those are the
main contents of Section II. In Section III, we will discuss the
entanglement between different DOF in a multi-photon state. It reveals the
entanglement between different DOF in the four-photon state from PDC, even
when there is no entanglement between them in the two-photon state. It is
the entanglement that makes decoherence in polarization DOF in the
experiments on the four-photon state. Our analysis is coincident with the
results of experiments in Ref. \cite{Tsujino,Eisenberg} which discussed the
coherence of polarization DOF of the four-photon state. The last section is
the conclusion.

\section{Photon state representation and two-photon interference}

Photon interference can be regarded as the result of QIP and can be well
described with the language of QIP. Usually, there is more than one DOF
involved in the multi-photon interference. Firstly, the photon system is a
boson, which satisfies the permutation symmetric principle. Naturally, the
state description should be permutation symmetric. From the photon creation
operators, the multi-photon state can be written as:

\begin{equation}
\dprod\limits_{i=1}^{N}a_{i}^{\dag }\left\vert vac\right\rangle =\frac{1}{%
\sqrt{N!}}\sum_{P}P(\left\vert a_{1}\right\rangle \left\vert
a_{2}\right\rangle ...\left\vert a_{N}\right\rangle )\text{,}
\label{nphoton}
\end{equation}%
where $P$ is the permutation of the $N$ single-photon states. When all the $%
a_{i}^{\dag }$ are the same, it degenerates to
\begin{equation}
a^{\dag N}\left\vert vac\right\rangle =\sqrt{N!}\left\vert a\right\rangle
^{\otimes N}\text{,}
\end{equation}%
where $\sqrt{N!}$ shows the photon bunching. Next, if there is no coupling
between different DOF, the single photon state is described as:
\begin{equation}
a^{\dag }(\alpha ,\beta ,...,\gamma )\left\vert vac\right\rangle =\left\vert
\alpha ,\beta ,...,\gamma \right\rangle =\left\vert \alpha \right\rangle
\left\vert \beta \right\rangle ...\left\vert \gamma \right\rangle \text{,}
\label{nDOF}
\end{equation}%
where $\alpha $, $\beta $, ... $\gamma $ are for different DOF. With Eqs.(%
\ref{nphoton}) and (\ref{nDOF}), any photon state can be described in a new
expression. Now, we will applly it to the two-photon state to re-interpret
the HOMI.

\begin{figure}[h]
\includegraphics[width=4cm]{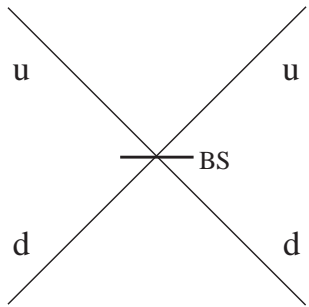}
\caption{Illustration of two-photon Hong-Ou-Mandel interference on the
Beamsplitter. The $u$ and $d$ describe the two different spatial mode.}
\label{fig1}
\end{figure}

The two-photon state involving different DOF is described as:%
\begin{eqnarray}
\left\vert \Phi _{2}\right\rangle  &=&c\sum_{\alpha _{i},\beta
_{i},...,\gamma _{i}}\phi (\alpha _{1},\beta _{1},...,\gamma _{1},\alpha
_{2},\beta _{2},...,\gamma _{2})  \notag \\
&&\times a^{\dag }(\alpha _{1},\beta _{1},...,\gamma _{1})a^{\dag }(\alpha
_{2},\beta _{2}...,\gamma _{2})\left\vert vac\right\rangle \text{,}
\end{eqnarray}%
where $c$ is the normalization number. In the usual cases, one DOF is path
mode and the other is polarization, frequency, time-energy, or transverse
spatial mode. Here, we consider the two-photon states from non-collinear
PDC. In this case, the two photons are in two different paths, which are
labeled as $u$ and $d$, as shown in Fig.\ref{fig1}. For simplicity, we here
only consider two DOF \cite{note}. Thus the state is written as%
\begin{eqnarray}
\left\vert \Phi _{2}\right\rangle  &=&c\sum_{\alpha _{1},\alpha _{2}}\phi
(\alpha _{1},u,\alpha _{2},d)a^{\dag }(\alpha _{1},u)a^{\dag }(\alpha
_{2},d)\left\vert vac\right\rangle   \notag \\
&=&\frac{c}{4\sqrt{2}}\sum_{\alpha _{1},\alpha _{2}}\{[\phi (\alpha
_{1},u,\alpha _{2},d)+\phi (\alpha _{2},u,\alpha _{1},d)](\left\vert \alpha
_{1},\alpha _{2}\right\rangle   \notag \\
&&+\left\vert \alpha _{2},\alpha _{1}\right\rangle )(\left\vert
u,d\right\rangle +\left\vert d,u\right\rangle )+[\phi (\alpha _{1},u,\alpha
_{2},d)  \notag \\
&&-\phi (\alpha _{2},u,\alpha _{1},d)](\left\vert \alpha _{1},\alpha
_{2}\right\rangle -\left\vert \alpha _{2},\alpha _{1}\right\rangle
)(\left\vert u,d\right\rangle -\left\vert d,u\right\rangle )\}\text{.}
\label{original}
\end{eqnarray}%
In the above expressions, $\alpha _{1}$ and $\alpha _{2}$ are in the same
basis. For convenience, we gather the notations which describe the same DOF.
From the above state, we will consider some special cases in the HOMI where
two photons are injected into the two input ports of the beamsplitter (BS)
simultaneously.

The lossless BS can be regarded as the transformation on the path DOF while
it has no effect on polarizations \cite{note2}. It is single qubit rotation
gate:%
\begin{eqnarray}
\left\vert u\right\rangle &\longrightarrow &\sqrt{R}\left\vert
u\right\rangle -\sqrt{T}\left\vert d\right\rangle )\text{,}  \notag \\
\left\vert d\right\rangle &\longrightarrow &\sqrt{T}\left\vert
u\right\rangle +\sqrt{R}\left\vert d\right\rangle )\text{,}
\end{eqnarray}%
where $R$\ and\ $T$\ are the reflectivity and transmissivity of BS with $%
R+T=1$.

For symmetric BS ($R=T=50\%$), the two photon state is transformed to
\begin{eqnarray}
\left\vert \Phi _{2}\right\rangle _{out} &=&\frac{c}{4\sqrt{2}}\sum_{\alpha
_{1},\alpha _{2}}\{[\phi (\alpha _{1},u,\alpha _{2},d)+\phi (\alpha
_{2},u,\alpha _{1},d)](\left\vert \alpha _{1},\alpha _{2}\right\rangle
\notag \\
&&+\left\vert \alpha _{2},\alpha _{1}\right\rangle )(\left\vert
u,u\right\rangle -\left\vert d,d\right\rangle )+[\phi (\alpha _{1},u,\alpha
_{2},d)  \notag \\
&&-\phi (\alpha _{2},u,\alpha _{1},d)](\left\vert \alpha _{1},\alpha
_{2}\right\rangle -\left\vert \alpha _{2},\alpha _{1}\right\rangle
)(\left\vert u,d\right\rangle -\left\vert d,u\right\rangle )\}\text{.}
\label{full}
\end{eqnarray}%
When $\phi (\alpha _{1},u,\alpha _{2},d)=\phi (\alpha _{2},u,\alpha _{1},d)$%
, the two photons are in the symmetric states, including the polarized Bell
triplet states
\begin{eqnarray}
\left\vert \Phi ^{\pm }\right\rangle  &=&\frac{1}{2}(\left\vert
HH\right\rangle \pm \left\vert VV\right\rangle )(\left\vert u,d\right\rangle
+\left\vert d,u\right\rangle )\text{,}  \notag \\
\left\vert \Psi ^{+}\right\rangle  &=&\frac{1}{2}(\left\vert HV\right\rangle
+\left\vert VH\right\rangle )(\left\vert u,d\right\rangle +\left\vert
d,u\right\rangle )\text{,}  \label{triplet}
\end{eqnarray}%
and superpositions of the three states. Photons in these states will cause
the photon bunching in the same output port of BS. If $\phi (\alpha
_{1},u,\alpha _{2},d)=-\phi (\alpha _{2},u,\alpha _{1},d)$, it is the
antisymmetric state or singlet state
\begin{equation}
\left\vert \Psi ^{-}\right\rangle =\frac{1}{2}(\left\vert HV\right\rangle
-\left\vert VH\right\rangle )(\left\vert u,d\right\rangle -\left\vert
d,u\right\rangle )\text{.}  \label{singlet}
\end{equation}%
Two photons will be in different path DOF after\ the BS. It shows the photon
anti-bunching in the output ports of the BS. The above deduction is also the
principle of partial discrimination of four Bell states with the BS.

From the above discussion, any single qubit transformation collectively on
one DOF of the particles does not change the symmetry of that DOF.
Especially for the anti-symmetric state $\left\vert \Psi ^{-}\right\rangle $%
, the state will still not change after passing through any other BSs with
arbitrary reflection and transmission. It is also valid for the polarization
DOF. Actually, any operation collectively on the polarization DOF of the two
photons state $\left\vert \Psi ^{-}\right\rangle $\ will not change the
state description for there is only one form in the antisymmetric subspace.
It is the rotation invariance property of the singlet state $\left\vert \Psi
^{-}\right\rangle $. That is:
\begin{equation}
\rho =U\otimes U\rho U^{\dag }\otimes U^{\dag }\text{,}
\end{equation}%
where $\rho =\left\vert \Psi ^{-}\right\rangle \left\langle \Psi
^{-}\right\vert $. Generally, the operations in operator sum representation
also keep the permutation symmetry. So the state $\left\vert \Psi
^{-}\right\rangle $ will still be unchanged under those operations and it is
the decoherence-free entangled state \cite{Zanardi}.

The BS also has no effect on the other DOF such as frequency, time-energy
and is still a single qubit gate on the path DOF. For those DOF, the system
can be multi-dimensional and can be dealt with similar to the polarization
DOF. After the HOMI, two photons will also show the photon bunching if $\phi
$ is permutation symmetric and show the photon anti-bunching if $\phi $ is
permutation antisymmetric.

However, it is more complicated for the BS that has effect on different DOF
simultaneously, such as the transverse spatial mode. For example, the HOMI
of the two-photon state containing polarization DOF and transverse spatial
mode DOF in Ref. \cite{Walborn} shows different results according to both
the symmetry of polarization DOF and the parity of the transverse spatial
mode DOF. It also can be successfully interpreted with the above method by
considering the BS as a two-qubit gate on transverse spatial mode DOF and
path DOF.

It is the HOMI that we must consider the permutation symmetry of the path
DOF. When the two paths never interact, it becomes simple, since photons can
be labeled with the path DOF. For example, Eq. (\ref{singlet}) is simplified
to normal form:%
\begin{equation}
\left\vert \Psi ^{-}\right\rangle =\frac{1}{\sqrt{2}}(\left\vert
H_{u}V_{d}\right\rangle -\left\vert V_{u}H_{d}\right\rangle )\text{,}
\end{equation}%
However, for the multi-photon state, especially the photons from process of
parametric down-conversion, there are more cases of more than one photon are
in one path DOF. It is not convenient to label each photon with path DOF.
The state description based on permutation symmetry is more natural.

Moreover, the state description reveals the correlations between different
photons and different DOF. It is not difficult to find that there is no
entanglement between the polarization DOF and the path DOF in Eq.(\ref%
{singlet}), while both polarization and path entanglement exists between two
photons. Along with Eq.(\ref{original}), when $\phi $ is symmetric or
antisymmetric, there is no entanglement between different DOF. For a single
DOF, it is a pure entangled state when other DOF are traced out. The two
photons can show perfect interference in either DOF. When the symmetry of $%
\phi $ is broken, there is entanglement between different DOF. The
decoherence will happen in one single DOF and the visibility of photon
interference will dropped if the other DOF are neglected. Especially in the
multi-photon system, this state description can conveniently be used to deal
with the entanglement and coherence between photons and DOF. Next, we will
show that entanglement exists between different DOF in the four-photon state.

\section{Entanglement between different DOF in the four-photon state}

As shown above, for the two photons from the PDC, even in the
hyper-entangled state \cite{Barreiro,Cinelli,Barbieri}, there is no evidence
that correlations exist between these DOF. However, there will be
entanglement between different DOF in the four-photon state from the PDC.

The state from the process of the PDC pumped by a coherent pulse can be
described as \cite{Ou}:%
\begin{equation}
\left\vert \Psi \right\rangle =(1-\eta ^{2}/2)\left\vert vac\right\rangle
+\eta \left\vert \Phi _{2}\right\rangle +\eta ^{2}\left\vert \Phi
_{4}\right\rangle +...\text{.}
\end{equation}%
$\left\vert \eta ^{2}\right\vert $ is the probability of the two-photon
conversion in a single pump pulse. The two-photon state $\left\vert \Phi
_{2}\right\rangle $ has the form of the Schmidt decomposition:
\begin{equation}
\left\vert \Phi _{2}\right\rangle =\sum_{i}\phi (\alpha _{i},\beta
_{i})a^{\dag }(\alpha _{i})b^{\dag }(\beta _{i})\left\vert vac\right\rangle
\text{.}
\end{equation}%
The real variables $\phi (\alpha _{i},\beta _{i})$ satisfy the normalization
condition $\sum_{i}\left\vert \phi (\alpha _{i},\beta _{i})\right\vert ^{2}=1
$. Correspondingly, the four-photon state is:%
\begin{eqnarray}
\left\vert \Phi _{4}\right\rangle  &=&\frac{1}{2}\sum_{i,m}\phi (\alpha
_{i},\beta _{i})\phi (\alpha _{m},\beta _{m})a^{\dag }(\alpha _{i})b^{\dag
}(\beta _{i})a^{\dag }(\alpha _{m})b^{\dag }(\beta _{m})  \notag \\
&&\times \left\vert vac\right\rangle \text{.}
\end{eqnarray}%
$\alpha _{i}$, $\beta _{i}$ denote the DOF of the photon, such as
polarization, spatial or temporal mode, etc.

Let us introduce a measurable coefficient $K$ which is defined as
\begin{equation}
K=\sum_{i}\phi ^{4}(\alpha _{i},\beta _{i})\text{.}
\end{equation}%
$K$ indicates the entanglement of the two-photon state. The less $K$, the
more entanglement between the two photons \cite{law}. It is not difficult to
get $K$ with the method of quantum state tomography. Also, $K$ can be
directly measured in experiment from the single-photon rate ($P_{1}$) and
two-photon rate ($P_{2}$) in one path mode: $K=P_{2}/P_{1}^{2}-1$ \cite%
{sun07}.

In details, we consider a hyper-entangled state from non-collinear PDC,
\begin{eqnarray}
\left\vert \Phi _{2}\right\rangle  &=&\sum_{i}\phi (a_{i},b_{i})\left\vert
a_{i},b_{i}\right\rangle \sum_{j}\varphi (\alpha _{j},\beta _{j})\left\vert
\alpha _{j},\beta _{j}\right\rangle   \notag \\
&=&\sum_{i,j}\phi (a_{i},b_{i})\varphi (\alpha _{j},\beta _{j})a_{u}^{\dag
}(a_{i},\alpha _{j})b_{d}^{\dag }(b_{i},\beta _{j})\left\vert
vac\right\rangle \text{.}
\end{eqnarray}%
where $a_{u}^{\dag }$ and $b_{d}^{\dag }$ are the photon creation operators
for two path modes $u$ and $d$. The italic letters ($a$, $b$) are for
different states of one DOF and the greek letters ($\alpha $, $\beta $) are
for the other DOF. There is no entanglement between two DOF in the
two-photon state. The four-photon from the second order PDC is
\begin{eqnarray}
\left\vert \Phi _{4}\right\rangle  &=&\frac{1}{2}\sum_{i,j,m,n}\phi
(a_{i},b_{i})\varphi (\alpha _{j},\beta _{j})\phi (a_{m},b_{m})\varphi
(\alpha _{n},\beta _{n})  \notag \\
&&\times a_{u}^{\dag }(a_{i},\alpha _{j})b_{d}^{\dag }(b_{i},\beta
_{j})a_{u}^{\dag }(a_{m},\alpha _{n})b_{d}^{\dag }(b_{m},\beta
_{n})\left\vert vac\right\rangle   \notag \\
&=&[\sqrt{K_{A}K_{B}}\left\vert A_{1}\right\rangle \left\vert
B_{1}\right\rangle +\sqrt{K_{A}(1-K_{B})/2}\left\vert A_{1}\right\rangle
\left\vert B_{2}\right\rangle   \notag \\
&&+\sqrt{(1-K_{A})K_{B}/2}\left\vert A_{2}\right\rangle \left\vert
B_{1}\right\rangle +\sqrt{(1-K_{A})(1-K_{B})}/2\left\vert A_{2}\right\rangle
\notag \\
&&\times \left\vert B_{2}\right\rangle +\sqrt{(1-K_{A})(1-K_{B})/4}%
\left\vert A_{3}\right\rangle \left\vert B_{3}\right\rangle ]/\sqrt{%
(1+K_{A}K_{B})/2}\text{,}
\end{eqnarray}%
where the bases are:%
\begin{eqnarray}
\left\vert A_{1}\right\rangle  &=&\frac{1}{\sqrt{K_{A}}}\sum_{i}\phi
^{2}(a_{i},b_{i})(\left\vert i,i,\right\rangle _{u}\left\vert
i,i,\right\rangle _{d})_{A}\text{,}  \notag \\
\left\vert B_{1}\right\rangle  &=&\frac{1}{\sqrt{K_{B}}}\sum_{j}\varphi
^{2}(\alpha _{j},\beta _{j})(\left\vert j,j\right\rangle _{u}\left\vert
j,j\right\rangle _{d})_{B}\text{,}  \notag \\
|A_{2}\rangle  &=&\frac{1}{\sqrt{(1-K_{A})/2}}\sum_{i<m}\phi
(a_{i},b_{i})\phi (a_{m},b_{m})[(\left\vert i,m\right\rangle   \notag \\
&&+\left\vert m,i\right\rangle )_{u}(\left\vert i,m\right\rangle +\left\vert
m,i\right\rangle )_{d}/2]_{A}\text{,}  \notag \\
|B_{2}\rangle  &=&\frac{1}{\sqrt{(1-K_{B})/2}}\sum_{j<n}\varphi (\alpha
_{j},\beta _{j})\varphi (\alpha _{n},\beta _{n}))[(\left\vert
j,n\right\rangle   \notag \\
&&+\left\vert n,j\right\rangle )_{u}(\left\vert j,n\right\rangle +\left\vert
n,j\right\rangle )_{d}/2]_{B}\text{,}  \notag \\
|A_{3}\rangle  &=&\frac{1}{\sqrt{(1-K_{A})/2}}\sum_{i<m}\phi
(a_{i},b_{i})\phi (a_{m},b_{m})[(\left\vert i,m\right\rangle   \notag \\
&&-\left\vert m,i\right\rangle )_{u}(\left\vert i,m\right\rangle -\left\vert
m,i\right\rangle )_{d}/2]_{A}\text{,}  \label{antisymmetricA} \\
|B_{3}\rangle  &=&\frac{1}{\sqrt{(1-K_{B})/2}}\sum_{j<n}\varphi (\alpha
_{j},\beta _{j})\varphi (\alpha _{n},\beta _{n}))[(\left\vert
j,n\right\rangle   \notag \\
&&-\left\vert n,j\right\rangle )_{u}(\left\vert j,n\right\rangle -\left\vert
n,j\right\rangle )_{d}/2]_{B}\text{.}  \label{antisymmetricB}
\end{eqnarray}%
The indices $i$, $m$ ($j$, $n$) are for the different photon states in $A$ ($%
B$) DOF. $K_{A}=\sum_{i}\phi ^{4}(a_{i},b_{i})$ and $K_{B}=\sum_{j}\varphi
^{4}(\alpha _{j},\beta _{j})$. So, $K_{A(B)}\leq 1$ can be measured with the
method mentioned above by breaking the entanglement with the DOF $B$ ($A$).
In the above four-photon state description, we neglect the permutation
symmetry between the different path DOF for they never meet and label
photons by $u$ and $d$. Also, we write the two-photon state with the same
path DOF in one ket and describe them in permutation symmetric or
anti-symmetric form.

\begin{figure}[h]
\includegraphics[width=5cm]{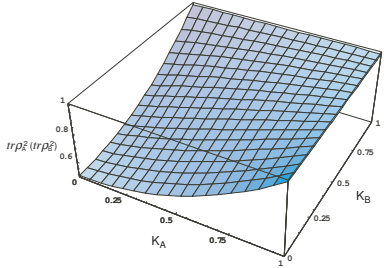}
\caption{(color online)Plot of the value of $tr\protect\rho _{A}^{2}$ ($tr%
\protect\rho _{B}^{2}$) with the varying of $K_{A}$ and $K_{B}$.}
\label{fig2}
\end{figure}

For simplicity, we make
\begin{eqnarray}
\left\vert A(B)_{12}\right\rangle  &=&(\sqrt{K_{A(B)}}\left\vert
A(B)_{1}\right\rangle   \notag \\
&&+\sqrt{(1-K_{A(B)})/2}\left\vert A(B)_{2}\right\rangle )/\sqrt{%
(1+K_{A(B)})/2}\text{.}  \label{symmetric}
\end{eqnarray}%
Then the four-photon state is:
\begin{eqnarray}
\left\vert \Phi _{4}\right\rangle  &=&(\sqrt{(1+K_{A})(1+K_{B})}\left\vert
A_{12}\right\rangle \left\vert B_{12}\right\rangle +\sqrt{(1-K_{A})(1-K_{B})}%
\left\vert A_{3}\right\rangle   \notag \\
&&\times \left\vert B_{3}\right\rangle )/\sqrt{2(1+K_{A}K_{B})}\text{.}
\label{four-photon-full}
\end{eqnarray}%
In Eq.(\ref{four-photon-full}), states $\left\vert A(B)_{12}\right\rangle $
and $\left\vert A(B)_{3}\right\rangle $ are the permutation symmetric and
antisymmetric forms for four photons in one DOF. The product keeps the
permutation symmetry of bosonic particles. It is easy to get
\begin{eqnarray}
\rho _{A} &=&tr_{B}\rho _{AB}=[(1+K_{A})(1+K_{B})\left\vert
A_{12}\right\rangle \left\langle A_{12}\right\vert +(1-K_{A})(1-K_{B})
\notag \\
&&\times \left\vert A_{3}\right\rangle \left\langle A_{3}\right\vert
]/(2+2K_{A}K_{B})\text{,}
\end{eqnarray}%
and
\begin{equation}
tr\rho _{A}^{2}=tr\rho _{B}^{2}=\frac{%
1+4K_{A}K_{B}+K_{A}^{2}+K_{B}^{2}+K_{A}^{2}K_{B}^{2}}{2(1+K_{A}K_{B})^{2}}%
\text{.}
\end{equation}%
When there is no entanglement in DOF $A$ ($B$) of the two-photon state, then
$K_{A(B)}=1$, $tr\rho _{A}^{2}=tr\rho _{B}^{2}=1$. This implies that there
is no entanglement between the two DOF in the four-photon state. In this
case, all photons are in the same state of DOF $A$ ($B$). The antisymmetric
part $\left\vert A_{3}\right\rangle $ ($\left\vert B_{3}\right\rangle $)
will not appear for it violates the permutation symmetry of photons.
However, if entanglement exists on both two single DOF of the two-photon
state, \textit{i.e.} hyper-entangled state, to make $K_{A(B)}\,<1$, the
antisymmetric part [Eqs.(\ref{antisymmetricA}) and (\ref{antisymmetricB})]
will appear. There is the entanglement between the two DOF in four-photon
state, although there is no entanglement between the two DOF in the
two-photon hyper-entangled state. As shown in Fig \ref{fig2}, when $%
K_{A}=K_{B}\rightarrow 0$, $tr\rho _{A}^{2}$ ($tr\rho _{B}^{2}$) approaches
the minimal value of $0.5$, which is the maximal entanglement between the
two modes.

As an example, we will discuss the system containing polarization DOF and
another DOF besides the path DOF. In Refs. \cite{Tsujino,Ou2,Eisenberg},
there exists other DOF entangling with the polarization DOF which makes
interference of the polarization state less than $100\%$. Here we briefly
interpret it by considering two DOF in the system. We also suppose there is
no entanglement between different DOF in the two-photon state for the high
visibility of two-photon interference. There are two reasons to induce the
result. One is that the polarization state is a mixed state as mentioned
above. The two-photon Einstein-Podolsky-Rosen state entangled in
polarization DOF is described as:%
\begin{equation}
\left\vert \Psi _{2}^{-}\right\rangle =(\left\vert H\right\rangle
_{u}\left\vert V\right\rangle _{d}-\left\vert V\right\rangle _{u}\left\vert
H\right\rangle _{d})/\sqrt{2}\text{.}
\end{equation}%
$K_{A}=1/2$, it is the maximally entangled state. If there is no other modes
correlated with polarization mode, then the four-photon state will be \cite%
{Tsujino}:
\begin{eqnarray}
\left\vert \Psi _{4}\right\rangle &=&(\left\vert HH\right\rangle
_{u}\left\vert VV\right\rangle _{d}-\left\vert HV\right\rangle
_{u}\left\vert HV\right\rangle _{d}+\left\vert VV\right\rangle
_{u}\left\vert HH\right\rangle _{d})/\sqrt{3}  \notag \\
&=&(\sqrt{2}\left\vert \pi _{1}\right\rangle +\left\vert \pi
_{2}\right\rangle )/\sqrt{3}\text{,}
\end{eqnarray}%
where $\left\vert \pi _{1}\right\rangle =\frac{1}{\sqrt{2}}(\left\vert
HH\right\rangle _{u}\left\vert VV\right\rangle _{d}+\left\vert
VV\right\rangle _{u}\left\vert HH\right\rangle _{d})$ and $\left\vert \pi
_{2}\right\rangle =-(\left\vert HV\right\rangle +\left\vert VH\right\rangle
)_{u}(\left\vert HV\right\rangle +\left\vert VH\right\rangle )_{d}/2$. $%
\left\vert \Psi _{4}\right\rangle $ is the symmetric state for polarization
DOF according to $\left\vert A_{12}\right\rangle $ in Eq.(\ref{symmetric}).
However, if there is another DOF, the two photon state will be $\left\vert
\Phi _{2}\right\rangle =\left\vert \Psi _{2}^{-}\right\rangle \otimes
\left\vert \Phi _{2}\right\rangle _{B}$. By tracing the mode $B$, the
four-photon state in the polarization DOF will be in the form of

\begin{equation}
\rho _{4}=[3(1+K_{B})\left\vert \Psi _{4}\right\rangle \left\langle \Psi
_{4}\right\vert +(1-K_{B})\left\vert \Phi _{4}\right\rangle \left\langle
\Phi _{4}\right\vert ]/(4+2K_{B})\text{,}  \label{four-photon}
\end{equation}%
where $\left\vert \Phi _{4}\right\rangle =(\left\vert HV\right\rangle
-\left\vert VH\right\rangle )_{u}(\left\vert HV\right\rangle -\left\vert
VH\right\rangle )_{d}/2$ is the anti-symmetric state for polarization DOF
according to $\left\vert A_{3}\right\rangle $ in Eq.(\ref{antisymmetricA}).
It is clear that the polarization DOF is entangled with the other DOF and
the four-photon polarization state is in a mixed state if $K_{B}\neq 1$.

The other reason comes from the measurement. On one path DOF, the two-photon
state measurement can be
\begin{eqnarray}
M &=&\sum_{i,j,m,n}f(a_{i},\alpha _{j})g(a_{m},\alpha _{n})  \notag \\
&&\times a^{\dag }(a_{i},\alpha _{j})a^{\dag }(a_{m},\alpha
_{n})a(a_{m},\alpha _{n})a(a_{i},\alpha _{j})\text{.}
\end{eqnarray}%
According to the two-photon cases, the measurement is rewritten as

\begin{eqnarray}
M_{u(d)} &=&\sum_{i,j,m,n}f(a_{i},\alpha _{j})g(a_{m},\alpha
_{n})[(\left\vert im\right\rangle +\left\vert mi\right\rangle
)_{A}(\left\vert jn\right\rangle +\left\vert nj\right\rangle )_{B}  \notag \\
&&+(\left\vert im\right\rangle -\left\vert mi\right\rangle )_{A}(\left\vert
jn\right\rangle -\left\vert nj\right\rangle _{B})][(\left\langle
im\right\vert +\left\langle mi\right\vert )_{A}(\left\langle jn\right\vert
\notag \\
&&+\left\langle nj\right\vert )_{B}+(\left\langle im\right\vert
-\left\langle mi\right\vert )_{A}(\left\langle jn\right\vert -\left\langle
nj\right\vert )_{B}]/8\text{.}
\end{eqnarray}%
The result of the measurement will be $P=tr(\rho _{AB}M_{u}\otimes M_{d})$.
Actually, the antisymmetric part will be invariant under any collective
operation and gives unchanged result. So the antisymmetric part of $%
\left\vert \Phi _{4}\right\rangle $ will contribute a "background" in the
measurement result if there is entanglement between the two DOF both in the
initial state and the measurement part. This will make the visibility of the
four-photon interference less than $100\%$ or induce the distinguishability
of the state.

In Refs.\cite{Tsujino,Eisenberg}, a four-photon polarization state is
described by the sum of two parts with a variable $\alpha $. It is a mixed
state as shown above in Eq.(\ref{four-photon}), or Eq.(5) in Ref. \cite%
{Eisenberg}. If we set $\alpha =3K_{B}/(2+K_{B})$, they describe the same
state. However, the two parts in the state described in Ref. \cite{Tsujino}
are not mutually orthogonal. For the polarization DOF correlated with the
frequency DOF case, it has been successfully discussed in the language of
multi-mode description \cite{Ou,Ou2,Sun2}. Comparing with Ref.\cite{Ou2}, we
find $K_{B}=\mathcal{E}/\mathcal{A}$. However, the descriptions in the above
references did not reveal the relations between different DOF while Eq.(\ref%
{four-photon-full}) gives a clear description of the existence of
entanglement between different DOF.

\section{Conclusion}

In conclusion, we described the photon interference with the language of
QIP. Together with the state expression based on the permutation symmetry of
boson nature, the two-photon HOMI is re-interpreted and the multi-photon
state involving different DOF is discussed. It reveals the existence of
entanglement between different DOF in the four-photon state even there is no
entanglement in the two-photon state. For a special case, the polarization
DOF entangled with other DOF is analyzed. It is the entanglement that makes
the decoherence in the polarization DOF in the experiment.

\begin{acknowledgments}
We thank Z. W. Zhou and X. F. Zhou for helpful discussion. This work was
funded by the Chinese National Fundamental Research Program, the Innovation
funds from Chinese Academy of Sciences, NCET, and National Natural Science
Foundation of China. ZYO is also supported by the US National Science
Foundation under Grant No. 0427647.
\end{acknowledgments}


\begin{thebibliography}{99}
\bibitem{Hong} C. K. Hong, Z. Y. Ou, and L. Mandel, Phys. Rev. Lett. \textbf{%
59} 2044 (1987).

\bibitem{Kok} P. Kok, W. J. Munro, K. Nemoto, T. C. Ralph, J. P. Dowling and
G. J. Milburn, Rev. Mod. Phys. \textbf{79}, 135 (2007).

\bibitem{Ou88} Z. Y. Ou and L. Mandel, Phys. Rev. Lett. \textbf{61}, 50
(1988).

\bibitem{Kiesel} N. Kiesel, C. Schmid, U. Weber, R. Ursin, and H.
Weinfurter, Phys. Rev. Lett. \textbf{95}, 210505 (2005).

\bibitem{Brien} J. L. O'Brien, G. J. Pryde, A. G. White, T. C. Ralph, and D.
Branning, Nature \textbf{426}, 264 (2003).

\bibitem{Braunstein} S. L. Braunstein and A. Mann, Phys. Rev. A \textbf{51},
R1727 (1995).

\bibitem{Mattle} K. Mattle, H. Weinfurter, P. G. Kwiat, and A. Zeilinger,
Phys. Rev. Lett. \textbf{76}, 4656 (1996).

\bibitem{Bouwmeester} D. Bouwmeester, J. W. Pan, K. Mattle, M. Eibl, H.
Weinfurter, and A. Zeilinger, Nature \textbf{390}, 575 (1997).

\bibitem{Ricci0} M. Ricci, F. De Martini, N. J. Cerf, R. Filip, J. Fiur\'{a}%
\v{s}ek, and C. Macchiavello, Phys. Rev. Lett. \textbf{93}, 170501 (2004).

\bibitem{Santori} C. Santori, D. Fattal, J. Vu\v{c}kovi\'{c}, G. S. Solomon,
and Y. Yamamoto, Nature \textbf{419} 594, 2002.

\bibitem{Ricci} M. Ricci, F. Sciarrino, C. Sias, and F. De Martini, Phys.
Rev. Lett. \textbf{92}, 047901 (2004)

\bibitem{Irvine} W. T. M. Irvine, A. L. Linares, M. J. A. de Dood, and D.
Bouwmeester, Phys. Rev. Lett. \textbf{92}, 047902 (2004).

\bibitem{Fonseca} E. J. S. Fonseca, C. H. Monken, and S. P\'{a}dua, Phys.
Rev. Lett. \textbf{82}, 2868 (1999).

\bibitem{D'Angelo} M. D'Angelo, M. V. Chekhova, and Y. Shih, Phys. Rev.
Lett. \textbf{87}, 013602 (2001).

\bibitem{Walther} P. Walther, J. -W, Pan, M. Aspelmeyer, R. Ursin, S.
Gasparoni, and A. Zeilinger, Nature(London) \textbf{429}, 158 (2004).

\bibitem{mitchell} M. W. Mitchell, J. S. Lundeen, and A. M. Steinberg,
Nature(London) \textbf{429}, 161 (2004).

\bibitem{sun1} F. W. Sun, B. H. Liu, Y. F. Huang, Z. Y. Ou, and G. C. Guo,
Phys. Rev. A \textbf{74}, 033812 (2006).

\bibitem{Vaziri} A. Vaziri, G. Weihs, and A. Zeilinger, Phys. Rev. Lett.
\textbf{89}, 240401 (2002).

\bibitem{Riedmatten} H. De Riedmatten, I. Marcikic, H. Zbinden, and N. Gisin
Quant. Inf. Comm. \textbf{2}, 425 (2002).

\bibitem{Mair} A. Mair, A. Vaziri, G. Weihs, and A. Zeilinger, Nature
(London) \textbf{412}, 313 (2001).

\bibitem{Marcikic} I. Marcikic, H. de Riedmatten, W. Tittel, H. Zbinden, and
N. Gisin, Nature (London) \textbf{421}, 509 (2003).

\bibitem{Barreiro} J. T. Barreiro, N. K. Langford, N. A. Peters, and P. G.
Kwiat, Phys. Rev. Lett. \textbf{95}, 260501 (2005).

\bibitem{Cinelli} C. Cinelli, M. Barbieri, R. Perris, P. Mataloni, and F. De
Martini, Phys. Rev. Lett. \textbf{95}, 240405 (2005)

\bibitem{Barbieri} M. Barbieri, C. Cinelli, P. Mataloni, and F. De Martini,
Phys. Rev. A \textbf{72}, 052110 (2005).

\bibitem{Tsujino} K. Tsujino, H. F. Hofmann, S. Takeuchi, and K. Sasaki,
Phys. Rev. Lett. \textbf{92}, 153602 (2004).

\bibitem{Eisenberg} H. S. Eisenberg, J. F. Hodelin, G. Khoury, and D.
Bouwmeester, Phys. Rev. Lett. \textbf{96}, 160404 (2006).

\bibitem{note} Actually, different DOF can be combined into one DOF in
mathematics just to span the basis.

\bibitem{note2} Here we neglect the extra $\pi $ phase shift when the
horizontally polarized photon is reflected by the BS. It can be compensated
by adding phase shifters on the outports of the BS.

\bibitem{Zanardi} P. Zanardi and M. Rasetti, Phys. Rev. Lett. \textbf{79},
3306 (1997).

\bibitem{Walborn} S. P. Walborn, A. N. de Oliveira, S. P\'{a}dua, and C. H.
Monken, Phys. Rev. Lett. \textbf{90}, 143601 (2003).

\bibitem{Ou} Z. Y. Ou, J. -K. Rhee, and L. J. Wang, Phys. Rev. A \textbf{60}%
, 593 (1999).

\bibitem{law} C. K. Law and J. H. Eberly, Phys. Rev. Lett. \textbf{92},
127903 (2004); K. Rza\.{z}ewski and J. H. Eberly, J. Phys. B \textbf{27},
L503 (1994). For consistency with $\mathcal{E}/\mathcal{A}$, we choose the
sum form but not the reciprocal form.

\bibitem{sun07} F. W. Sun, \textit{et. al, }Phys. Rev. A \textbf{76}, 052303
(2007).

\bibitem{Ou2} Z. Y. Ou, Phys. Rev. A \textbf{72}, 053814 (2005).

\bibitem{Sun2} F. W. Sun, Z. Y. Ou, and G. C. Guo, Phys. Rev. A \textbf{73},
023808 (2006).
\end{thebibliography}
\end{document}